\documentclass[conference]{IEEEtran}
\IEEEoverridecommandlockouts
\usepackage{cite}
\usepackage{amsmath,amssymb,amsfonts}
\usepackage{algorithmic}
\usepackage{graphicx}
\usepackage{textcomp}
\usepackage{xcolor}
\usepackage{footnote}
\usepackage{makecell}
\usepackage{tikz}
\usetikzlibrary{shapes,arrows}
\def\BibTeX{{\rm B\kern-.05em{\sc i\kern-.025em b}\kern-.08em
    T\kern-.1667em\lower.7ex\hbox{E}\kern-.125emX}}

\definecolor{bleudefrance}{rgb}{0.19, 0.55, 0.91}
\definecolor{chromeyellow}{rgb}{1.0, 0.65, 0.0}
\definecolor{asparagus}{rgb}{0.53, 0.66, 0.42}

\begin{document}
\title{Simulation Framework for Cooperative Adaptive Cruise Control with Empirical DSRC Module\\
}


\author{\IEEEauthorblockN{1\textsuperscript{st} Zijia Zhong} 
\IEEEauthorblockA{\textit{Department of Civil and Environmental Engineering} \\
\textit{University of Delaware}\\
Newark, DE, USA \\
zzhong@udel.edu}
\and
\IEEEauthorblockN{2\textsuperscript{nd} Joyoung Lee}
\IEEEauthorblockA{\textit{Department of Civil and Environmental Engineering} \\
\textit{New Jersey Institute of Technology}\\
Newark, NJ, USA \\
jo.y.lee@njit.edu}
}

\maketitle
\begin{abstract}
Wireless communication plays a vital role in the promising performance of connected and automated vehicle (CAV) technology. This paper proposes a Vissim-based microscopic traffic simulation framework with an analytical dedicated short-range communication (DSRC) module for    packet reception. Being derived from ns-2, a packet-level network simulator, the DSRC probability module takes into account the imperfect wireless communication that occurs in real-world deployment.  Four managed lane deployment strategies are evaluated using the proposed framework.  While the average packet reception rate is above 93\% among all tested scenarios, the results reveal that the reliability of the vehicle-to-vehicle (V2V) communication can be influenced by the deployment strategies. Additionally, the proposed framework exhibits desirable scalability for traffic simulation and it is able to evaluate transportation-network-level deployment strategies in the near future for CAV technologies.
\end{abstract}

\begin{IEEEkeywords}
Dedicated short-range communication, imperfect wireless communication, cooperative adaptive cruise control, microscopic traffic simulation, automated driving system fallback
\end{IEEEkeywords}

\section{Introduction}
Cooperative adaptive cruise control (CACC), as one of the most promising connected and automated vehicle (CAV) technologies, is expected to revolutionize the way vehicles are operated, making the modern transportation systems faster, safer and greener. CACC is enabled by the dedicated short-range communication (DSRC), which is designed for vehicle-to-vehicle (V2V) communication and vehicle-to-infrastructure (V2I) communication. (V2V and V2I are often jointly referred as V2X.) The term "Dedicated" refers to the 75-MHz licensed spectrum in the 5.9 GHz band assigned by the U. S. Federal Communications Commission (FCC). The term “"short-range"” indicates that the communication taking places is only over hundreds of meters.  Wireless access in vehicular environment (WAVE)\cite{xiang2009wireless} message is broadcasted via DSRC for safety-critical application, whereas the non-safety-critical application can use either DSRC protocol stack or other wireless communication protocols. Wireless communication in vehicular environment is a technological challenge due to: 1) limited communication channels are accessed by a multitude of communication nodes (vehicles); 2) the aforementioned communication nodes are commonly travelling at high speed (e.g., 120 km/h); 3) the metallic exterior of a vehicle increases the reflection of radio, resulting a more complicated communication environment. 

Large-scale field experiments on CACC on actual roadway are prohibitive and the safety is also the primary concerns at the current stage of CAV. Simulation study is still one of the most cost-effective ways to assess the impacts of CACC. Previous state-of-the-art research has been focusing on the vehicle dynamic aspect of CACC, studying the mechanical aspect of CACC manipulations, for instance, how the acceleration was achieved by analyzing each subcomponent (e.g., throttle position, sensor gain, vehicle powertrain). Such aspect is vitally important in bringing CACC into fruition; yet, it provides little insights into the impacts of CACC on the overall traffic and transportation network. In addition, studies of CACC's impacts on traffic from simulation thus far have been mostly done with the assumption of perfect wireless communication. The performance of wireless communication has seldom studied on the operational aspect of CACC deployment on large-scale network yet. 
	
This paper focuses on examining the impact of DSRC in near-term CACC deployment. A state-of-the-art traffic simulation framework, which incorporates both a traffic simulator and an analytical module for packet reception is developed. The remainder of this paper is organized as follows. Section \ref{LitRvw} summarizes the relevant research conducted previously, followed by the Methodology in Section \ref{SecMethod}. The simulation scenario design and results are provided in Section \ref{SecRslt}. Lastly, conclusions for this paper are offered in Section \ref{SecCon}. 
\newline

\section{Literature Review} \label{LitRvw}
There are primarily three types of simulations for CACC evaluation. In the transportation engineering side, simulations focus on high-level vehicular interactions. A traffic simulator (TS) with a realistic car-following model is of the center of the simulation. In traffic simulation, the majority of the studies has been conducted under the assumption of perfect V2X communication. Lee \textit{et al.} \cite{jiang2007communication} evaluated the mobility and safety impact of CACC using Vissim with an active platooning algorithm. Arnaout and Bowling \cite{arnaout2014progressive} evaluated three different deployment strategies for CACC with a custom simulation testbed. The benefits of CACC start to show when the market penetration rate (MPR) reaches 40\% in the absence of any managed lane strategies.  In \cite{shladover2012recent}, a microscopic TS was developed using the Aimsun and its microSDK package \cite{lu2014freeway} to assess the improvement of highway capacity by ACC and CACC. From the electrical engineering standpoint, a packet-level network simulator (NS) is typically used.  Akhtar \textit{et al.} \cite{akhtar2015vehicle} used MATLAB for implementing channel model to post-process the vehicle trajectory data obtained by SUMO, an open-source TS. Other research using a packet-level NS can be also found in \cite{chen2009ieee,elbatt2006cooperative,xu2004vehicle}. All of these studies, however, shared the inadequacy of using primitive traffic behavioral models that may not reflect the realistic vehicular behaviors in the real world.  

Efforts in combining NSs and TSs have also been reported.  The majority of the efforts attempted to couple a NS and a TS by using an external synchronizing module. SUMO \cite{behrisch2011sumo} and OMNeT++ \cite{varga2010omnet} were combined by a synchronizer via TCP/IP connection in \cite{ucar2016security,motro2016vehicular, segata2014plexe}. While being able to conduct realistic traffic simulation and communication simulation simultaneously, the frameworks were suffered from poor scalability: no more than 10 vehicles were tested during the simulation for the three aforementioned studies. Veeraraghavan and Miloslavov \cite{miloslavov2012integrated} synchronized NCTUns \cite{wang2010nctuns} and Vissim \cite{ptv2015vissim} to conduct CAV traffic simulation to the tune of 300 vehicles per hour (vph). Eichler \textit{et al.} \cite{eichler2005simulation} coupled ns-2, an empirical packet-level network simulator,  and CARIMA \cite{holly1993description} and conducted simulations with 400 equipped vehicles in an 8-km$^{2}$ city area.  All of these research provides a realistic communication simulation, but it also lacks the scalability for successful implementation of thousands of CAVs per hour that passing through typical thoroughfare in the real world. 

	To scale up, hybrid simulation is believed to be a viable option, which can achieve a significant reduction in the number of scheduled events through the use of analytical models while maintaining the credibility of simulation.  As in most scenarios, the key question is whether a subject vehicle successfully receives messages from another vehicle. The rest of the data traffic can be treated as background data traffic.  Hence, the transmission scenarios are simplified during simulation runtime.  Jiang \textit{et al}. put forward the concept of communication density, which serves as a metric for channel load in vehicular communication. The communication density is defined as the number of sensible events per unit of time and it is the production of vehicle density, message generation rate, and communication range \cite{jiang2007communication}. Data reception rate is determined at a particular communication density level with a particular transmission power.  Channel assess delay is an important aspect in vehicle safety communication and it is defined as the duration between the arrival of a frame at the medium access control (MAC) layer to the point of transmission over the air.  Jiang \textit{et al.} \cite{jiang2007communication} proved that average channel access stays the same in the cases where communication density levels are the same. Further developing from Jiang \textit{et al.}\textquotesingle s concept of communication density level, Killat and Hartenstein \cite{killat2009empirical} proposed an analytical model derived from ns-2-based 5.9GHz DSRC simulation. Levenberg-Marquardt method \cite{weisstein2000levenberg} was used to construct a two-dimensional polynomial curve fitting. The model assumed 382-byte packet size (128 bytes for the certificate, 54 bytes for signature, and 200 bytes of available payload). Within the highest data transmission rate of 27 Mbit/s, the maximum communication density can be handled is 4400 in theory. The probability of one-hop broadcast packet reception in WAVE can be computed based on the above assumption.   

Among the previous studies, we found studies only focus on the packet-level communication simulation with no or unrealistic vehicle movements. Other studies post-processed the communication simulation, which means the communication has no impact on vehicle movements. The third group of studies used a synchronizer between a TS and a NS, trying to gain the advantages for both types of simulators. But the simulation speed suffered greatly when data is being exchanged between a TS and a NS.  Hence, a hybrid microscopic simulation framework is desired, which can implement CACC vehicle behaviors, provide realistic vehicular interaction, and at the same time factor in the communication impacts at a much larger scale. 
\newline


\section{Methodology} \label{SecMethod}
\subsection{Probability Model for DSRC}
Killat \textit{et al.} \cite{killat2007enabling} have proven that the number of scheduled events in ns-2 drastically increases as the traffic density increase within a network, resulting in time-consuming simulation and poor scalability. Hence, it is computationally unsound to conduct a packet-level network assessment with thousands of vehicles, a typical scale which traffic engineers are accustomed to for traffic impact analysis. As such, Killat and Hartenstein \cite{killat2009empirical} proposed an analytical model built from ns-2. The model is developed upon Jiang \textit{et al.}\textquotesingle s concept of communication density level, which is a metric of representing channel load in vehicular communication in the form of the sensible transmission per unit of time and per unit of the road\cite{jiang2007communication}. 

	Killat’s model yields the probability of one-hop broadcast reception under DSRC. For a single sender, the model is a combination of Nakagami m-distribution fast fading model and the Friis/TRG(two-ray-ground) path loss model \cite{simon2005digital}. For multiple senders, a statistical model is derived from scenarios of a single sender with the Levernberg-Marquardt method.  Compared to pure ns-2-based approach, the hybrid approach is more computationally tractable-500 speedup factor in a network of 2500 to 3000 vehicles-with similar characteristics of transmission simulated by ns-2. The probability of a successful packet reception by a receiving vehicle can be calculated as \eqref{eq:recpProb} and \eqref{eq:commDen}

\begin{gather}
P_{r}(x, \delta , \varphi , f)=e^{^{-3(x/\varphi )^{2}}}(1+ \sum_{i=1}^{4}h_{i}(\xi ,\varphi )(\frac{x}{\varphi})^{i})
\label{eq:recpProb}
\end{gather}
where, ${h_{i}(\xi ,\varphi)}$ is the two-dimensional polynomial of fourth-degree for all curving fitting parameters; $\xi$ is communication density, events/s/km; and $\varphi$ is the transmission power, m.

\begin{gather}
\xi = \delta  \cdot \varphi \cdot f
\label{eq:commDen}
\end{gather}
where, $\delta $ is vehicle per kilometer that periodically broadcast messages, veh/km; $f$ is transmission rate, Hz; and $\xi, \varphi$ are as previously defined.

\begin{figure} [!ht]
	\centering
	\includegraphics[width=\columnwidth]{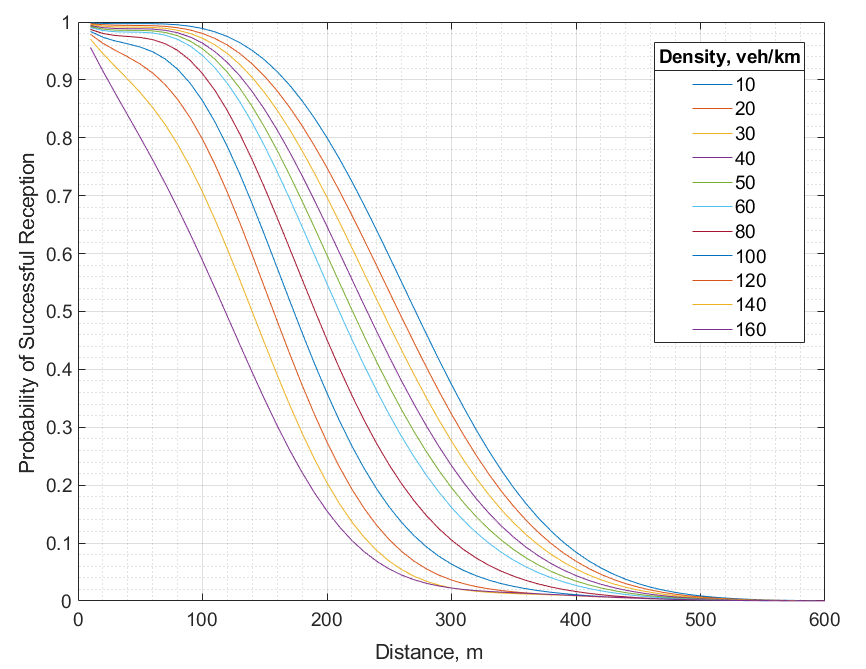}   
	\caption{Probability density curves for the analytical model} 
	\label{fig:dsrcPdf}
\end{figure}

Fig. \ref{fig:dsrcPdf} shows the probability density curve of the DSRC wireless communication model under different communication density, with the power range of 300 m \cite{shladover2015cooperative} and the transmission frequency of 10 Hz \cite{ngoduy2013instability}. It is worth pointing out the assumptions made for the wireless communication module:  1) the channel access is assumed zero. 2) high data rate in this simulation has the same level of countering noise and interference of the low data rate.
\newline

\subsection{Simulation Framework}
\begin{figure}[tbp]
\centering
\resizebox{\columnwidth}{!}
{%
\tikzstyle{decision} = [trapezium, draw, trapezium right angle =70, trapezium left angle = 110, minimum width =3cm, minimum height = 1cm, fill=yellow, text centered, text width = 3cm, rounded corners]
\tikzstyle{block} = [rectangle, draw, fill=bleudefrance, 
    text width=5em, text centered, rounded corners, minimum height =4em, minimum width =8em]
\tikzstyle{end} = [rectangle, draw, fill=asparagus, 
    text width=5em, text centered, rounded corners, , minimum height =4em]
\tikzstyle{line} = [draw, -latex']

\begin{tikzpicture}[node distance = 2cm, auto]
    \node [block,  text width = 10em] (killatModel) {Killat's Nakagami-based analytical model};
    \node [block, above of =killatModel, text width = 10em] (getData) {communication parameters};
    \node [block, below of=killatModel, text width = 10em] (calProb) {compute reception probability};
    \node [block, below of=calProb, text width = 10em] (ranNum) {random number test};
    \node [decision, below of=ranNum] (experiment) {random number $<$ reception probability};
    \node [block, below of=experiment, xshift=-3cm, node distance=2cm, text width=8em] (success) {V2V communication successful};
    \node [block, below of=experiment, xshift=3cm, node distance=2cm, text width=8em] (fail) {instance of packet drop, instance++};
    \node [decision, above of=fail, text width=6em, yshift=2cm, xshift=2cm] (attemptChek) {total attempts $<$ 5};
    \node [end, below of=success , text width=10em ] (shortHw) {use the short following headway};
    \node [end, below of=fail, text width=10em] (longHw) {use the long following headway};
    \path [line] (getData) -- (killatModel);
    \path [line] (killatModel) -- (calProb);
    \path [line] (calProb) -- (ranNum);
    \path [line] (ranNum) -- (experiment);
    \path [line] (experiment) -| node [anchor = east, yshift = 0.3cm, xshift = 0.7cm] {yes} (success);
    \path [line] (success) -- (shortHw);
    \path [line] (experiment) -| node [anchor = west, yshift = 0.3cm, xshift = -0.7cm]{no}(fail);
    \path [line] (fail) -|([xshift=0.5cm]fail.east) -- (attemptChek);
    \path [line] (attemptChek)|-node[anchor = east, yshift = -3cm]{yes}(getData);
    \path [line] (attemptChek)-|([xshift=2.5cm]fail.south east) |- node[anchor = east, yshift = 6.2cm]{no}(longHw);
\end{tikzpicture}
}
\caption{DSRC packet reception testing procedure} 
\label{fig:testPrced}
\end{figure}
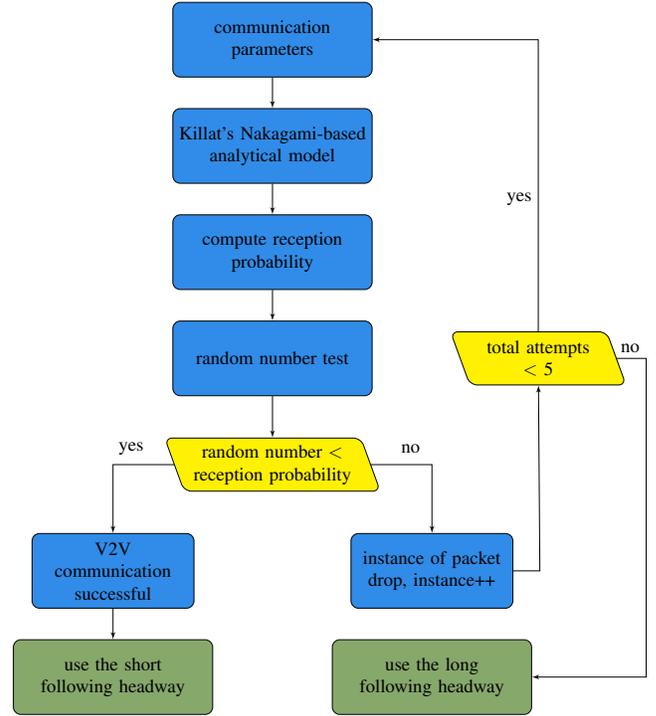

The proposed simulation framework is comprised of Vissim, Vissim COM interface (VCOM) \cite{vissim2015introduction}, Vissim DriverModel.DLL (VEDM)\cite{ptv2015vissim}. Vissim is a multimodal microscopic traffic simulation software in which each entity (e.g., car, train, pedestrian) is simulated individually. The VEDM allows users to replace the Wiedemann car-following model of Vissim with a custom one. In this study, the multi-objective-optimization (MOOP)-based platoon maneuvering framework \cite{zhong2017multiobjective} is used for the longitudinal control of each CACC vehicle, which is comprised of four objectives-mobility, safety, emission, and fuel consumption-to be optimized. It is assumed that, for each simulation time step, a CACC vehicle has up to five attempts to transmit the vehicle status message. Each attempt is considered as an independent experiment. If the latest packet is not delivered, a CACC vehicle uses the second latest packet to conduct the optimization. Hence, as long as a packet is successfully transmitted within the five attempts, the wireless transmission is deemed successfully for communicating vehicle status. The reception probability is computed for each platooned CACC vehicle based on \eqref{eq:recpProb}. A random number is generated within the range of [0, 1]. If the random number is less than the current reception probability (e.g., 0.93), a transmission attempt is considered successful. The overall procedure for the packet reception experiment is illustrated in Fig. \ref{fig:testPrced}.

\pagestyle{empty}

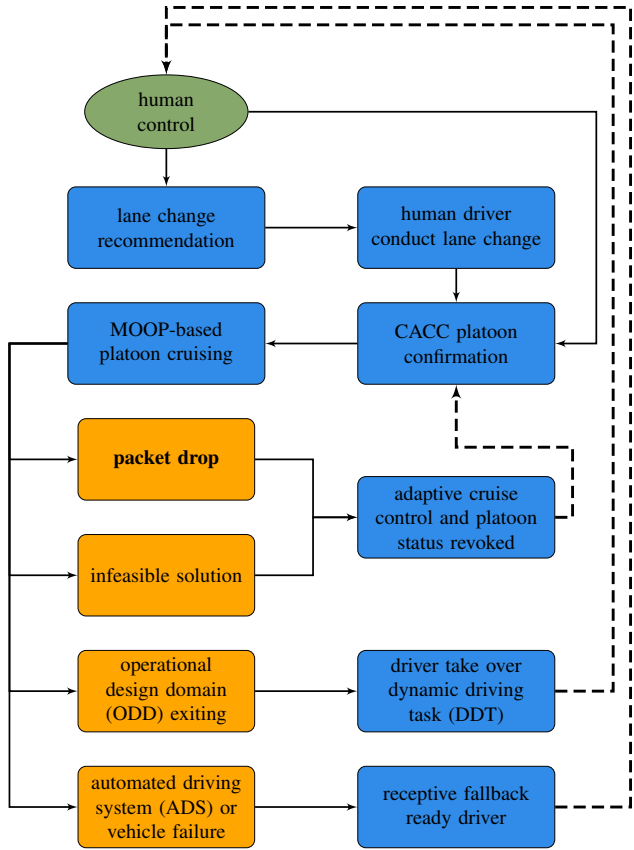
\begin{figure}[tbp]
\centering
\resizebox{\columnwidth}{!}
{%
\tikzstyle{block} = [rectangle, draw, fill=bleudefrance, 
    text width=9em, text centered, rounded corners, , minimum height =4em]
\tikzstyle{event} = [rectangle, draw, fill=chromeyellow, 
    text width=8em, text centered, rounded corners, minimum height =4em]
\tikzstyle{human} = [ellipse, draw, fill=asparagus, 
    text width=5em, text centered]
\tikzstyle{line} = [draw, -latex', line width = 0.3mm,]
\tikzstyle{fbkLine} = [draw, thick, color=black, -latex', dashed, line width = 0.5mm, dash pattern=on 6pt off 3pt]

\begin{tikzpicture}[node distance = 2cm, auto]
    \node[human, xshift = -4cm](start){human control};
    \node[block, below of =start, node distance = 2cm](laneChange){lane change recommendation};
    \node[block, right of =laneChange, xshift = 3cm](conductLc){human driver conduct lane change};
    \node[block, below of = conductLc](confirmation){CACC platoon confirmation};
    \node[block, left of = confirmation, , xshift = -3cm] (cruising){MOOP-based platoon cruising};
    \node[event, below of =cruising](pktDrop){\textbf{packet drop}};
    \node[event, below of =pktDrop](infeasible){infeasible solution};
    \node[event, below of =infeasible](OddExit){operational design domain (ODD) exiting};
    \node[event, below of =OddExit](ADSFail){automated driving system (ADS) or vehicle failure};
    \node[block, right of =infeasible, yshift=1cm, node distance = 5cm](ACCmode){adaptive cruise control and platoon status revoked};
    \node[block, right of =OddExit, node distance = 5cm](takeover){driver take over dynamic driving task (DDT)};
    \node[block, right of =ADSFail, node distance =5cm](readyDri){receptive fallback ready driver};
    \path [line] (start) -- (laneChange);
    \path [line] (laneChange) -- (conductLc);
    \path [line] (conductLc) -- (confirmation);
    \path [line] (confirmation) -- (cruising);
    \path [line] (cruising) -| ([xshift=-1cm,yshift=-0.5cm]cruising.south west)|-(pktDrop);
    \path [line] (cruising) -| ([xshift=-1cm,yshift=-0.5cm]cruising.south west)|-(infeasible);
    \path [line] (cruising) -| ([xshift=-1cm,yshift=-0.5cm]cruising.south west)|-(OddExit);
    \path [line] (cruising) -| ([xshift=-1cm,yshift=-0.5cm]cruising.south west)|-(ADSFail);
    \path [line] (pktDrop) -|([xshift=1cm]pktDrop.east)|-(ACCmode);
    \path [line] (infeasible) -|([xshift=1cm]infeasible.east)|-(ACCmode);
    \path [line] (OddExit)--(takeover);
    \path [line] (ADSFail)--(readyDri);
    \path [fbkLine] (takeover) -| ([xshift=1cm]takeover.east)|- ([xshift=5cm, yshift=1.5cm]start.east)-| (start);
    \path [fbkLine] (readyDri) -| ([xshift=1.3cm]readyDri.east)|-  ([xshift=5cm, yshift=1.8cm]start.east)-|(start);
    \path [fbkLine] (ACCmode) -| ([xshift=0.3cm]ACCmode.east)|- ([yshift =0.5cm]ACCmode.north)-|(confirmation);
    \path [line] (start) -| ([xshift=6cm]start.east)|- (confirmation);
\end{tikzpicture}
}
\caption{CACC system fallback} 
\label{fig:cavFallback}
\end{figure}

The Society of Automotive Engineer International (SAE) has defined five levels of vehicular automation as well as the corresponding system fallback in \cite{sae2016taxonomy}. CACC is classified as the SAE Level 3 automation that requires a receptive fallback ready driver. The overall fallback procedure is illustrated in Fig. \ref{fig:cavFallback}. The failure of packet delivery is counted as one of the four types of events that requires system fallback. When packet delivery is unsuccessful, reversion to adaptive cruise control with the longer time headway is performed. Due to the scope of the paper, only packet drop and the occurrence of infeasible MOOP solution are considered in the simulation.
\newline

\section{Simulation and Results} \label{SecRslt}
\subsection{Simulation Design}
\begin{figure} [!ht]
	\centering
	\includegraphics[width=80mm]{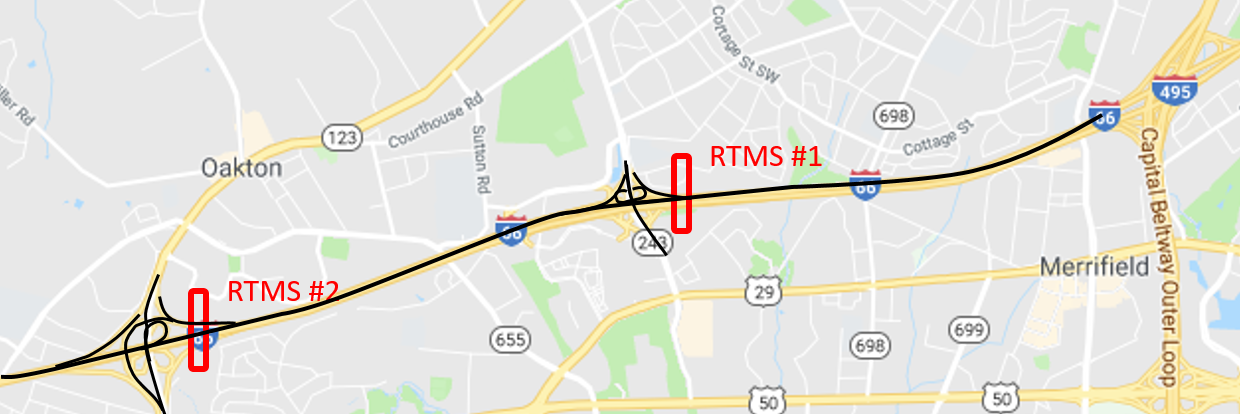}   
	\caption{Interstate highway 66 testbed} 
	\label{fig:i66Map}
\end{figure}

An 8-km (5-mile) segment of Interstate Highway I-66 outside of the beltway (I-495) of Washington D.C. (shown in Fig. \ref{fig:i66Map}) is selected as the freeway testbed. The roadway is a major commuter corridor with four lanes in each direction. The segment has recurring congestion during weekdays, specifically in the morning in the eastbound direction and in the afternoon in the westbound direction. The leftmost lane is a high-occupancy vehicle (HOV) lane, and the HOV lane has no physical barrier preventing access to/from the adjacent lane. The current demand for the network during PM-peak in the westbound direction is approximately 6000 vph.

Network calibration has been conducted by using two independent data sources-INRIX travel time data and volume data collected by remote traffic microwave sensors (RTMSs).  There are two RTMSs located before the exits of the freeway as marked in Fig. \ref{fig:i66Map}. The impact of lane use in terms of the reliability of DSRC was investigated based on the following scenarios under various MPRs: 

\begin{itemize}
\item{\textbf{Base case (BASE)}: This scenario serves as the base condition of the I-66 segment for this study. As stated, an HOV lane is implemented in the leftmost lane and the network has been calibrated.}
\item{\textbf{Unmanaged lane (UML)}: In this strategy, the HOV lane is revoked, and current HOV vehicles are treated as GP vehicles. CACC vehicles are not given priority use of any lanes, and they operate along with general purpose (GP) vehicles/human-driven vehicles.}
\item{\textbf{Dedicated CACC lane (DL)}: The exclusive access to the leftmost lane for CACC vehicles is studied. A homogeneous CACC traffic is believed to be beneficial for the CACC operation. The merging impact of CACC vehicles to the leftmost lane can be studied as well. Like in UML, the HOV vehicles are treated as GP vehicles in this strategy.}
\item {\textbf{Dedicated CACC lane with Access Control (DLA)}: This strategy is essentially a dedicated CACC lane mentioned above but with access control where a CACC vehicle is only able to merge in/out of the managed lane at designated locations. Therefore, the weaving activities are aggregated at certain locations of the network. It is formulated to insulate the CACC platoons from the potential impacts of weaving activities. }
\end{itemize}
A break-down list of the managed lane strategy is shown in TABLE \ref{table: MLs}. The fourth lane is the leftmost lane and the initial of each vehicle type is used.

\begin{table}[!h]
\centering
\caption{Scenario settings.} 
\begin{tabular}{cccccc}
\hline
Strategy & 4th & 3rd & 2nd & 1st & Access Control \\ \hline
BASE & H & G & G & G & N \\
UML & G+C  & G+C & G+C & G+C & N\\
MML & C+H & G & G & G & N\\
DL & C & G & G & G & N\\
DLA & C & G & G & G & Y\\
\hline
\end{tabular}
\label{table: MLs}
\end{table}

The simulation period for one replication is set as 3900 seconds with 300 seconds of a warm-up period to saturate the network with traffic. For each combination of managed lane strategy and MPR, five random replications are run to capture the variability of the traffic flow. The simulation resolution is set as two, which means the vehicle trajectory is calculated for every 0.5s. The optimization is conducted in every five simulation time steps.  The simulation is run with the following assumptions: 

\begin{itemize}
\item{The calibrated vehicle behaviors in Vissim realistically represent the road users’ driving behaviors.}
\item{The CACC controller is free of control errors.}
\item{The lateral control for platoon formation is conducted by human drivers with recommendations for lane change from the CACC system.}
\item{Human-driven vehicles treat CACC vehicles as human-driven vehicles. There are no indications whether a vehicle is equipped with CACC system.}
\end{itemize}

\subsection{Simulation Results}
\begin{figure} [h]
	\centering
	\includegraphics[width=0.95\columnwidth]{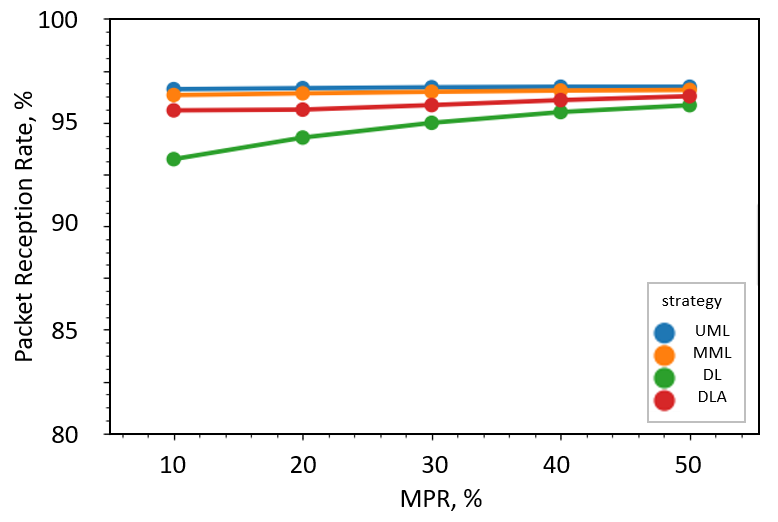}   
	\caption{Packet reception rate} 
	\label{fig:packRecp}
\end{figure}

The mean probability of successful packet reception is shown in Fig. \ref{fig:packRecp}. While the reception probability is above 0.9 for all managed lane cases, DL has the lowest probability among all strategies at any given MPR.  Since the V2V communication only occurs for CACC vehicles, under low MPR, CACC vehicles are scattered throughout the network, which could result in longer transmission distance. The probability of successful reception increases as more CACC vehicles get introduced to the network.  In the case without dedicated CACC lane (i.e., UMN and MML), the distributions of mean transmission density are similar.  DLA has the lowest median and variance among all four strategies at each MPR.

\begin{figure} [h]
	\centering
	\includegraphics[width=\columnwidth]{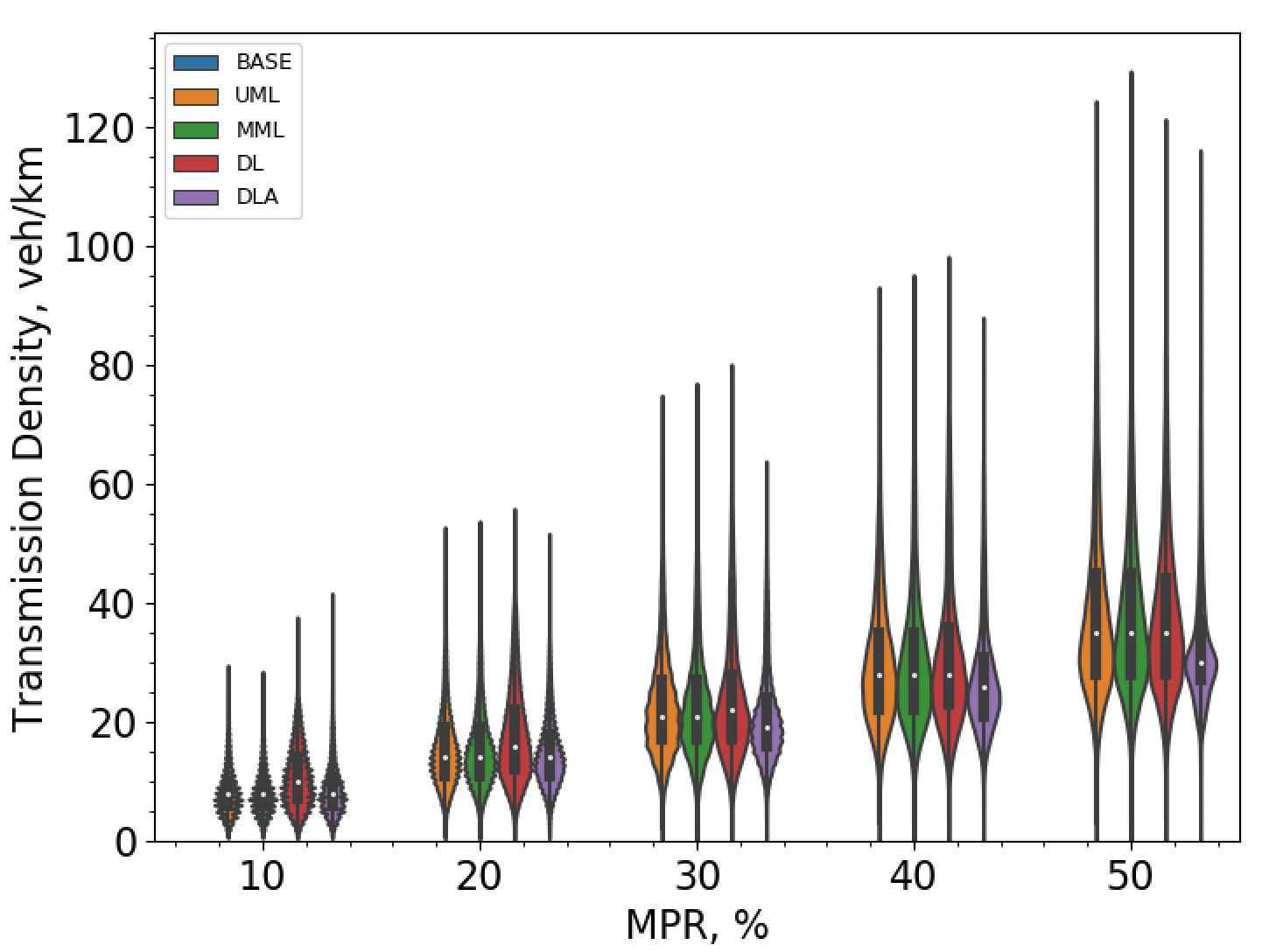}   
	\caption{Communication density} 
	\label{fig:commDens}
\end{figure}

The DL has the lowest reception rate at any given MPR.  At MPR below 30 \%, relatively higher transmission than the remaining three strategies was observed as shown in Fig. \ref{fig:commDens}. The Violin plot of DL shows greater variance on the transmission density. The DL strategy does not have any physical barriers, without which CACC vehicles are able to get onto the dedicated lane at any location of the roadway.  When a slower CACC vehicle merges into the CACC lane, faster CACC platoons behind it may need to slow down in order to maintain a safe following distance.  Under such scenario, the traffic string is compressed, yielding closer following distance and higher vehicle density. When limited access control to the dedicated lane is implemented, the disturbance to the CACC vehicles that are cruising on the dedicated lane is minimized, effectively reducing the events of deceleration as a result of cut-ins of slow-moving vehicles. Therefore, the transmission density maintains at a lower level with less variance, which is beneficial for the packet delivery.
\newline

\section{Conclusions} \label{SecCon}

This paper incorporates a one-hop DSRC probability model for CACC into Vissim, a microscopic traffic simulation software. By only considering the channel load of the wireless network, the scalability of the simulation for CAV technology is greatly increased, to the scale of thousands of vehicles, which is typical and necessary for traffic impact analysis. The proposed framework has been successfully tested on a large-scale, real-world transportation network with the demand of approximately 6000 vph. The impact of data packet delivery is considered during the simulation, which is one of the major advantages of the proposed framework. This study also evaluates the performance of V2V communication under the four managed lane strategies. From the perspective of wireless communication, the higher the packet reception probability, the more reliable communication among platooned CACC vehicles. The influence of managed lane strategy on the transmission density, and ultimately communication reliability is studied as well. The diminishing return of the packet reception ratio may not sometimes be justifiable among other factors (e.g., platoon size, total platoons, or the percentage of platooned CACC vehicles). However, the reliability of wireless communication has to be weighed in, when the packet reception ratio of certain managed lane strategies is out of the acceptable range.  The managed lane strategies may remedy the communication reliability by controlling the transmission density within a certain acceptable range. As demonstrated, the impact of wireless communication is crucial and should be integrated to the traffic simulation to present a more comprehensive picture of the near-term deployment of CACC.
\newline

\appendix
\subsection{The coefficients of $h^{(j,k)}_{i}$} \label{FirstAppendix}
The coefficients obtained from the polynomial function $h_{i}(\xi ,\varphi )$ is shown in TABLE \ref{table: hCoef}. It is worth stressed that even seemingly negligible values, if omitted, could result deviation in the probability of reception from 8\% to 100\%\cite{killat2009empirical}.
\begin{table}[!h]
\centering
\caption{Coefficient $h^{(j,k)}_{i}$ in \eqref{eq:recpProb} \cite{killat2009empirical}}
\resizebox{\columnwidth}{!}{%
\begin{tabular}{ccccccc}
\hline
 &  &  & $(j,k)$  &  &  \\
\hline
 &(0,0) & (1,0) & (2,0) & (3,0) & (4,0)  \\ 
\hline
$h^{(j,k)}_{1}$ & 0.0209865 & -9.66304$e$-07  &-1.72786$e$-11  & 5.09506$e$-17  & -7.91921$e$-23 \\
$h^{(j,k)}_{2}$ & 2.24743 & 7.84884$e$-07 & 2.28533$e$-10 & -5.89802$e$-16 & 3.55262$e$-22 \\
$h^{(j,k)}_{3}$ & 2.56426 & 2.82287$e$-05 & -7.09939$e$-10  & 1.34371$e$-15 & -3.01956$e$-22\\
$h^{(j,k)}_{4}$ & 2.41146 & -9.32859$e$05 & 6.77403$e$-10 & -9.64188$e$-16 & 3.69652$e$-23\\ 
\hline
 & (3,1) & (2,1) & (2,2) & (1,1) & (1,2)\\
\hline
$h^{(j,k)}_{1}$ & 3.16577$e$-20 & 2.13587$e$-14 & -5.05716$e$-17 & 4.00928$e$-09 & -1.88707$e$-11 \\
$h^{(j,k)}_{2}$ & 4.07120$e$-19 & -2.66510$e$-13 & 8.64273$e$-17 & -7.31274$e$-08 & 2.98549$e$-10 \\
$h^{(j,k)}_{3}$ & -1.85451$e$-18 & 1.02847$e$-12 & 1.80250$e$-16 & 1.56259$e$-07 & -8.50944$e$-10 \\
$h^{(j,k)}_{4}$ & 1.85043$e$-18 & -1.13894$e$-16 & -4.05333$e$-16 & -2.56738$e$-08 & 6.24415$e$-10 \\
\hline
 & (1,3) & (0,1) & (0,2) & (0,3) & (0,4)\\
\hline
$h^{(j,k)}_{1}$ & 3.25406$e$-14 & 0.000418109 & -4.30875$e$-06 & 1.00775$e$-08 & -7.32254$e$-12 \\
$h^{(j,k)}_{2}$ & -3.24982$e$-13 & 0.00498750 & -7.22232$e$-06 & 1.69755$e$-08 & -2.94381$e$-11\\
$h^{(j,k)}_{3}$ & 7.59094$e$-13 & -0.0227008 & 7.50391$e$-05 & -1.81469$e$-07 & 2.02182$e$-10\\
$h^{(j,k)}_{4}$ &-3.57571$e$-13 & 0.0191490 & -6.92678$e$-07 & 1.79917$e$-07 & -2.07263$e$-10\\
\hline
\end{tabular}
}
\label{table: hCoef}
\end{table}

\newpage
\bibliographystyle{IEEEtran}
\bibliography{IECON_DSRC}
\end{document}